\documentclass[final,1p,times]{elsarticle}

\usepackage{amssymb}

\journal{Nuclear Physics A}

\begin{document}

\begin{frontmatter}



\title{Properties of hyperons in chiral perturbation theory}


\author{J. Martin Camalich$^1$, L.S. Geng$^1$, L. Alvarez-Ruso$^2$, M.J. Vicente Vacas$^1$}

\address{$^1$Departamento de F\'{\i}sica Te\'orica and IFIC, Universidad de
Valencia-CSIC, Spain\\
$^2$Departamento de F\'isica, Universidade de Coimbra, Portugal}

\begin{abstract}
The development of chiral perturbation theory in hyperon phenomenology has been troubled due to power-counting subtleties
and to a possible slow convergence. Furthermore, the presence of 
baryon-resonances, e.g. the lowest-lying decuplet, complicates the approach, and the inclusion of their effects may become necessary. Recently, we have shown that a fairly good convergence is possible using a 
renormalization prescription of the loop-divergencies which recovers 
the power counting, is covariant and 
consistent with analyticity. Moreover, we have systematically incorporated the decuplet resonances taking care of both power-counting 
and $consistency$ problems. 
A model-independent understanding of diferent properties including the magnetic moments of the 
baryon-octet, the electromagnetic structure of the decuplet resonances and the 
hyperon vector coupling $f_1(0)$, has been successfully achieved within this approach. We will briefly review these developments and stress 
the important role they play for an accurate determination of the 
Cabibbo-Kobayashi-Maskawa matrix element $V_{us}$  from hyperon semileptonic decay data.
\end{abstract}


\end{frontmatter}


\section{Magnetic Moments}
\label{MMs}

The magnetic moments of the baryons are of the utmost importance since they contain information on their internal structure as read out by 
electromagnetic probes. A starting point is the SU(3)$_F$-symmetric model of Coleman and Glashow (CG)~\cite{Coleman:1961jn} that describes 
baryon-octet magnetic moments in terms of two parameters. The success of this model relies on the almost 
preserved global SU(3)$_V$-symmetry of QCD with $u$, $d$ and $s$ flavors. The description of the symmetry-breaking corrections of the baryon magnetic moments
can be addressed in a systematic and model-independent fashion by means of chiral perturbation theory ($\chi$PT)
~\cite{Gasser:1984gg,Gasser:1987rb}. In this approach, the CG result appears naturally at leading-order (LO) as tree-level. 
At next-to-leading (NLO) order, there are only loop-contributions that depend on known couplings and masses and, therefore,
no new undetermined low-energy constants (LECs) besides those appearing in the CG approach are to be included. The question is then if 
the SU(3)$_F$-breaking corrections to the baryon-octet magnetic moments can be successfully addressed from a first principles approach
by means of $\chi$PT; namely whether or not the chiral loops improve the classical CG results. A positive answer to this question has 
been given only recently~\cite{Geng:2008mf,Geng:2009hh} when applying the extended-on-mass-shell (EOMS) renormalization scheme ~\cite{Fuchs:2003qc}
which is an extension of $\overline{MS}$ developed to overcome the power-counting problem in the baryon sector of $\chi$PT. For a detailed 
presentation of our results and their comparison with heavy baryon (HB)~\cite{Jenkins:1990jv,Jenkins:1992pi}
or infrared (IR)~\cite{Becher:1999he,Meissner:1997hn} formulations of baryon $\chi$PT, see
Refs.~\cite{Geng:2008mf,Geng:2009hh}. The comparison of our results with those obtained before stresses the importance in SU(3)$_F$-$\chi$PT of the relativistic 
corrections, in the case of HB, and of keeping unaltered the analytic properties of the theory, in the case of IR. Concerning the  
inclusion of the decuplet resonances, $natural$ contributions that do not spoil the improvement over CG were found only in the EOMS
framework and when the unphysical degrees of freedom contained in the relativistic spin-3/2 vector-spinor were removed by means of the
$consistent$ couplings~\cite{Pascalutsa:2000kd}. It is also noteworthy that we obtain a good convergence since the NLO contribution is, 
at most, about one half of the LO one, what is consistent with our $a\;priori$ maximal expectation of $\sim m_\eta/\Lambda_{\chi SB}$.  

The aforementioned covariant approach that includes both octet and decuplet contributions has also been 
applied to the description of the electromagnetic structure of the decuplet resonances~\cite{Geng:2009ys}. In particular, the
magnetic dipole moments of the $\Delta^+$ and $\Delta^{++}$ are predicted using the well-measured one of the $\Omega^-$ to
fix the only LEC appearing up to NLO
\begin{equation}
  \mu_{\Delta^{++}}= 6.0(6)\mu_N,\hspace{1cm}\mu_{\Delta^+}=2.84(34)\mu_N, 
\end{equation}
where the error bars are an estimation of higher-order contributions obtained looking at the ratio between NLO and LO contributions
(we take $30\%$ of the NLO over LO ratio)~\cite{Geng:2009ys}. 
The relevance of these results lies on the ongoing efforts from the experimental side to measure the magnetic moments of these two 
resonances~\cite{Machavariani:1999fr,Kotulla:2002cg,Kotulla:2008zz}. On the theoretical side,
calculations from many different approaches have arisen in the last decades~\cite{Geng:2009ys}. Our results are 
compatible with the values quoted
by the Particle Data Group~\cite{Amsler:2008zzb} and the agreement with the latest experimental analysis, 
$\mu_{\Delta^{++}}=6.14\pm0.51\mu_N$~\cite{LopezCastro:2000cv}, is excellent. 

\section{Hyperon vector coupling $f_1(0)$}

The Cabibbo-Kobayashi-Maskawa (CKM) matrix ~\cite{Cabibbo:1963yz,Kobayashi:1973fv}
plays a very important role in our study and understanding of flavor
physics. In particular, its low mass sector allows for a precise test of the Standard Model through the CKM unitarity relation,
\begin{equation}
|V_{ud}|^2+|V_{us}|^2+|V_{ub}|^2=1,\label{Unitarity}
\end{equation}
 where one needs accurate values for $V_{ud}$, $V_{us}$, and $V_{ub}$.
Among them, $V_{ub}$ is quite small and can be
neglected at the present precision. The element  $V_{ud}$ can be obtained from
superallowed nuclear beta, neutron and pion decays, whereas
$V_{us}$ can be extracted from kaon, hyperon, and tau
decays (for a recent review, see Ref.~\cite{Amsler:2008zzb}). We now focus on how to determine $V_{us}$ from hyperon semileptonic 
decay data. 

The hyperon matrix elements of the weak flavor-changing currents are described by three vector (axial) form factors $f_i(q^2)$ ($g_i(q^2)$)
with $i=1,2,3$. The decay ratio of the semileptonic decay $B\rightarrow b l\bar{\nu}$ will then be determined by 
these form factors, the Fermi constant $G_F$, and the CKM element $V_{us}$. Indeed, if we define as a relevant SU(3)-breaking parameter
$\beta=\frac{M_B-M_b}{M_B}$, we can perform a power expansion of the decay rate about the SU(3)-symmetric limit
\begin{eqnarray}
&R\sim G_F^2 V_{us}^2 \left(\left(1-\frac{3}{2}\beta+\frac{6}{7}\beta^2\right)f_1^2+\frac{4}{7}\beta^2f_2^2+
\left(3-\frac{9}{2}\beta+\frac{12}{7}\beta^2\right)g_1^2+\frac{12}{7}g_2^2+\right.\nonumber\\
&\left.\frac{6}{7}\beta^2f_1f_2+\left(-4\beta+6\beta^2\right)g_1g_2 +\mathcal{O}(\beta^3)\right)\label{Eq:Rat3}
\end{eqnarray}
where the form factors are evaluated at $q^2=0$, although a linear $q^2$ dependence in $f_1$ and $g_1$
must also be considered at this order~\cite{FloresMendieta:2004sk}. Moreover, the SU(3)-symmetric limit for 
$f_2$ can be used. The most relevant contributions to the ratio come then from $g_1$, $f_1$ and
also $g_2$. Therefore, in order to extract accurately $V_{us}$ from semileptonic hyperon decay data, one requires to understand, 
in a model-independent fashion, the SU(3)-breaking contributions to these moments. The $g_2$ vanishes in the SU(3)-symmetric limit,
and we will assume $g_2=0$. The axial charge $g_1$, which is described in the 
symmetric limit by the parameters $D$ and $F$, receives $\mathcal{O}(\beta)$ breaking corrections. Nevertheless, as it has been proposed in 
Ref.~\cite{Cabibbo:2003cu}, we can use the measured $g_1/f_1$ ratios as the basic experimental data to equate $g_1$ in terms of $f_1$ in 
Eq.~(\ref{Eq:Rat3}). On the other hand, $f_1$ is protected by the Ademollo-Gatto Theorem~\cite{Ademollo:1964sr} which 
states that  breaking corrections start at $\mathcal{O}(\beta^2)$. 

\begin{table}
\caption{Results on the relative SU(3)-breaking of $f_1(0)$ in $\%$ for different channels obtained in $\chi$PT up to NNLO including octet and decuplet
contributions and those obtained in other approaches.\label{Table}}
\hspace{1cm}
\begin{tabular}{c|cccccc}
&B$\chi$PT&HB$\chi$PT&Large $N_c$&QM&$\chi$QM&lQCD\\
\hline\hline
$\Lambda \;N$&$+0.1^{+1.3}_{-1.0}$&+5.8&$+2\pm2$&$-1.3$&+0.1&\\
$\Sigma\;N$&$+8.7^{+4.2}_{-3.1}$&+9.3&$+4\pm3$&$-1.3$&+0.9&$-1.2\pm2.9\pm4$\\
$\Xi\Lambda$&$+4.0^{+2.8}_{-2.1}$&+8.4&$+4\pm3$&$-1.3$&+2.2&\\
$\Xi\Sigma$&$+1.7^{+2.2}_{-1.6}$&+2.6&$+8\pm5$&$-1.3$&+4.2&$-1.3\pm1.9$\\
\hline
\end{tabular}
\end{table}

The Ademollo-Gatto theorem is a consequence of the underlying SU(3)$_V$ symmetry of QCD, which has also important consequences 
when addressing a calculation of $f_1(0)$ in $\chi$PT. Namely, one finds that no unknown LECs contributing to this vector charge
are allowed until chiral order $\mathcal{O}(p^5)$. Therefore, a loop calculation up to and including NNLO
only depends on known masses and couplings and is a genuine prediction of $\chi$PT. Moreover, there are not divergencies 
or power counting breaking terms up to this order so that a counting restoration procedure does not seem necessary in this case.
This program has been developed in different steps along the last two decades~\cite{Anderson:1993as,Kaiser:2001yc,Krause:1990xc,
Lacour:2007wm,Villadoro:2006nj,Geng:2009ik}. A full NNLO calculation including
both octet and decuplet contributions in the covariant framework has been undertaken recently~\cite{Geng:2009ik}. In the latter work 
the problem with the convergence found in the HB calculation of Ref.~\cite{Villadoro:2006nj} has also been explained and fixed. 

In Table~\ref{Table} we present the results for the relative SU(3)-breaking correction $\frac{1}{100}\left(\frac{f_1(0)}{f_1^{SU(3)}(0)}-1\right)$ 
 in covariant $\chi$PT (B$\chi$PT) and HB$\chi$PT including octet and decuplet contributions up to NNLO. 
We also present those obtained in Large $N_c$~\cite{FloresMendieta:2004sk}, in a quark 
model (QM)~\cite{Donoghue:1986th}, in a chiral quark model ($\chi$QM)~\cite{Faessler:2008ix} and in lattice QCD
~\cite{Guadagnoli:2006gj,Sasaki:2008ha}. The error bars in the B$\chi$PT are an estimation of higher order uncertainties~\cite{Geng:2009ik}.
The results quoted
from  Ref.~\cite{Donoghue:1986th} are quite general in quark model calculations and reflect the 
naive expectation that  SU(3)-breaking corrections, at least for the $\Sigma N$ channel, should be negative. On the other hand, 
the different chiral approaches agree in the positive sign and the approximate size of these corrections, what may indicate 
the non-triviality of the multiquark effects induced by the chiral dynamics. It is also remarkable the agreement with
those obtained in a different systematic approach to non-perturbative QCD as the Large $N_c$. The results of lattice QCD are 
marginally compatible with ours although they favor negative corrections to $f_1(0)$. However, it must be pointed out that the pion masses 
in these simulations are still rather high, namely $\sim$400 MeV for $\Xi^0\rightarrow\Sigma^+$~\cite{Sasaki:2008ha} and $\sim$700 MeV for $\Sigma^-\rightarrow
n$~\cite{Guadagnoli:2006gj}. Another issue to be highlighted is the chiral extrapolation of the lattice QCD results to the physical point,
for which our results might be helpful in the future. And the other way around, the lattice QCD could provide information about the higher-order
local contributions in the chiral approach and could reduce the theoretical uncertainty of the B$\chi$PT calculation~\cite{Guadagnoli:2006gj}. 
In any case, a lattice simulation close to the physical point will be very helpful and eventually conclusive about the nature of the SU(3)-breaking
corrections to $f_1(0)$. 

With the elements developed above we obtain a determination of the CKM element $V_{us}$  that $combines$ the information on 
the different channels and includes 
the experimental errors~\cite{Mateu:2005wi} and higher-order errors estimated for $f_1(0)$ in B$\chi$PT
~\cite{Geng:2009ik}
\begin{equation}
V_{us}=0.2176\pm0.0029\pm\Delta_V, \label{vus}
\end{equation}
where $\Delta_V$ accounts for other systematic uncertainties. At the order we work in Eq.~(\ref{Eq:Rat3}), this uncertainty is due to
the SU(3)-breaking correction to $g_2$ that has not been considered. This contribution is $\sim\mathcal{O}(\beta^2)$  and
potentially as important for the extraction of $V_{us}$ as the SU(3)-breaking correction to $f_1$. 

We first compare our result with other determinations obtained from the decay rates and
$g_1/f_1$ in the hyperon semileptonic data; namely,  $V_{us}=0.2199\pm0.0026$ in Large $N_c$~\cite{FloresMendieta:2004sk}  and
$V_{us}=0.2250(27)$ in the SU(3)-symmetric model~\cite{Cabibbo:2003cu}. The comparison with the latter indicates the sensitivity to a 
breaking correction to $f_1(0)$ of $\sim\mathcal{O}(\beta^2)$ and suggests that the
SU(3)-symmetric assumption is not reliable enough for the accuracy required by the determination of $V_{us}$. The agreement between the B$\chi$PT
and the Large $N_c$ is a consequence of the consistency shown in Table~\ref{Table} and of the fact that in both approaches the SU(3)-breaking correction
to $g_2$ have been ignored.

On the other hand, our result is somewhat smaller than the ones obtained from kaon and tau decays or from the $f_K/f_\pi$ ratio~\cite{Amsler:2008zzb}.
It is not compatible either with the unitarity condition Eq. (\ref{Unitarity}) when using the value obtained from superallowed 
beta decays~\cite{Amsler:2008zzb}. Nonetheless, the result shown in Eq. (~\ref{vus}) is not complete and has to be improved with the
model-independent description of the SU(3)-breaking corrections to $g_2$. As argued in Ref.~\cite{Cabibbo:2003cu}, the trends shown by $\Sigma^-
\rightarrow n$ and $\Lambda\rightarrow p$ data indicate that the incorporation of the SU(3)-breaking corrections to $g_2$ will raise the value of
$V_{us}$ in these two channels.  Unfortunately, the data for hyperon decays is not yet precise enough to address a quantitative study of this 
form factor. From the theoretical side, a determination of these corrections in lattice QCD and an analysis in B$\chi$PT 
would be useful to ascertain the effects that $g_2$ may have on the determination of $V_{us}$.



\end{document}